\documentclass[aps,floats,floatfix,showpacs,amssymb,prl,twocolumn,groupedaddress,nofootinbib,nolongbibliography,10pt]{revtex4-2}

\usepackage{amssymb,amsmath,verbatim,mathtools,needspace,enumitem,etoolbox,graphicx,physics,microtype,afterpage,bigints,textcomp,gensymb,tabularx,bm,booktabs,multirow,ragged2e,nicefrac,colortbl,scalerel}

\usepackage[dvipsnames]{xcolor}
\definecolor{linkcolor}{rgb}{0.0,0.3,0.5}
\definecolor{cosmiclatte}{HTML}{FFF8E7}
\usepackage[unicode, colorlinks=true, linkcolor=linkcolor, citecolor=linkcolor, filecolor=linkcolor,urlcolor=linkcolor, pdfusetitle]{hyperref}
\usepackage[all]{hypcap}
\usepackage[T1]{fontenc}
\usepackage[utf8]{inputenc}
\usepackage{orcidlink}

\newcommand{\ssim}{\mathchar"5218\relax\,}

\makeatletter
\newcommand*{\balancecolsandclearpage}{\close@column@grid \cleardoublepage \twocolumngrid}
\makeatother

\usepackage{lipsum}

\newcommand{\jhu}{\affiliation{William H. Miller III Department of Physics and Astronomy,\\ Johns Hopkins University, 3400 North Charles Street, Baltimore, Maryland, 21218, USA}}

\begin{document}

\title{What does the Universe sound like?}

\author{Francesco Iacovelli\texorpdfstring{\,}{ }\orcidlink{0000-0002-4875-5862}}\email{fiacovelli@jhu.edu}
\jhu


\date{April 1, 2026}

\begin{abstract}
  Unlike electromagnetic telescopes, gravitational-wave (GW) detectors cannot produce pretty pictures, but we can convert GW signals into sound. I compute what the Universe actually sounds like by averaging over $\sim10^6$ synthetic compact binary coalescence events occurring throughout 2026. The result: a soothing, low-frequency rumble, perfect for sleeping, meditation, or contemplating the violent nature of spacetime. This is the \emph{Universal harmony}, audio file included!
\end{abstract}

\maketitle

\section{Introduction}

In the last decade, more than 200 gravitational-wave (GW) signals from compact binary coalescences (CBCs) have been observed~\cite{LIGOScientific:2025slb}. This has helped the field of GW astronomy to blossom. These observations at the LIGO, Virgo, and KAGRA detectors~\cite{LIGOScientific:2014pky,VIRGO:2014yos,Aso:2013eba} give us insight into the demography of binaries of compact objects across the Universe~\cite{KAGRA:2021duu,LIGOScientific:2025pvj}. They also reveal the nature of the signal-producing objects~\cite{LIGOScientific:2021sio,LIGOScientific:2025rid}, their internal structure~\cite{LIGOScientific:2017vwq}, and information about the Universe through which they travel~\cite{LIGOScientific:2021aug,LIGOScientific:2025jau}. 

Despite their tremendous potential, there is one thing GW detectors cannot provide compared to electromagnetic telescopes: pretty pictures. The wavelength of the gravitational radiation emitted by a CBC is typically greater than the size of the source. Hence, GWs cannot form an image; they can be best thought of as sound. One might then be left to wonder: what sound? And more importantly, is it any good?

For CBCs, the signal emitted by a single system increases in amplitude and frequency as the binary moves closer and closer to merger, resulting in a chirp-like sound~\cite{Maggiore:2007ulw}, similar to the one birds make. The exact frequency of the signal depends on the masses of the objects composing the binary: massive binaries, made up of black holes (BHs) of $10^6\,{\rm M}_\odot$, produce sounds in the milli-Hz frequency range or even lower, which is too low for the human ear. These are the Universe's bassists, playing notes so deep that only future space-based detectors like LISA~\cite{LISA:2017pwj,LISA:2024hlh} will appreciate their contribution to the cosmic symphony.
Binary black holes (BBHs) with components of a few to hundreds of solar masses, observable with ground-based detectors such as LIGO, Virgo, and KAGRA, are instead in the frequency range audible to the human ear, $f\gtrsim20\,$Hz. As an example, the sound produced by the very first GW detection, GW150914~\cite{LIGOScientific:2016wkq}, and the highest signal-to-noise ratio detection to date, GW250114~\cite{LIGOScientific:2025rid}, can be heard at Ref.~\cite{LVK250114video}. 

Current GW signal detections occur every few days. However, we expect more than $10^5$ CBC events per year in the Universe~\cite{LIGOScientific:2025pvj}--potentially several per minute. These signals can last for a few minutes in the audible band. So, how does the Universe sound? How do these chirps combine? Do they harmonize, or is it more of a cacophony? This is the soundtrack of the Universe, the rhythm at which we all vibrate, and what I will provide in this paper. Consider this the Universe's debut album, billions of years in the making.

\section{Methodology}

Let me begin by stressing that, by construction, this analysis will completely depend on the choices I make for the population of CBCs in the Universe. Nonetheless, my aim is to lay the foundations for a methodology that could be applied to any catalog of synthetic GW events and produce a Universal soundtrack for a representative, physically motivated population choice. I am simply choosing which instruments to include in the cosmic orchestra.

\begin{figure}[t]
    \centering
    \includegraphics[width=\linewidth]{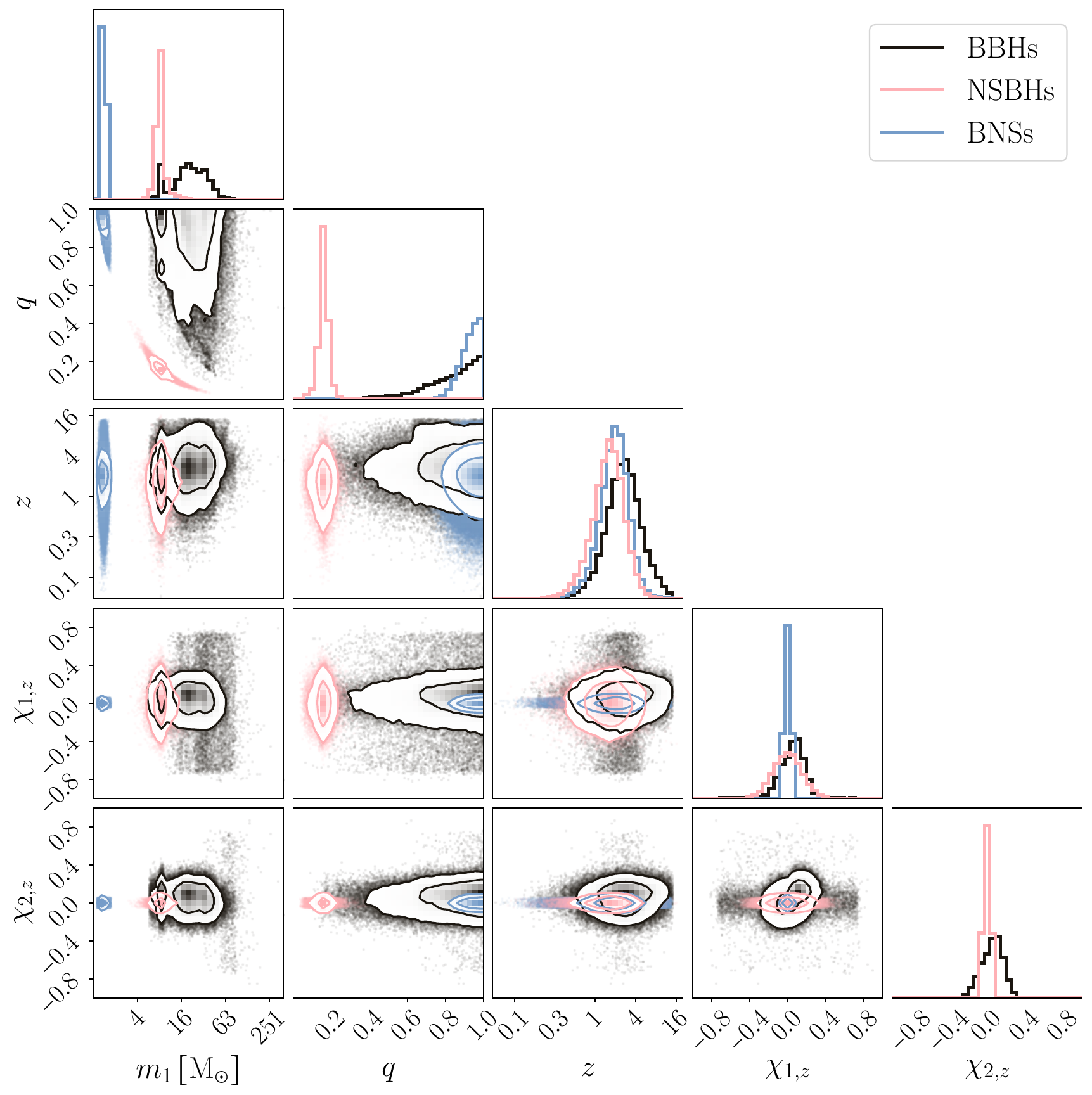}
    \caption{Distribution of the primary source-frame masses, mass ratios, redshifts, and spins of the chosen BBH, NSBH and BNS populations.  These parameters determine the pitch, timbre, and duration of each cosmic chirp.}
    \label{fig:universal_harmony_corner}
\end{figure}

I select the 1\,yr catalogs of BBH and binary neutron star (BNS) events employed in Ref.~\cite{Branchesi:2023mws}, produced from state-of-the-art population synthesis codes. These catalogs consist of $\ssim1.2\times10^5$ and $\ssim7.2\times10^5$ CBC events, respectively. For neutron star-black hole (NSBH) systems, instead, I use the same catalog as in Ref.~\cite{Iacovelli:2022bbs}, consisting of $\sim4.5\times10^4$ events. The distributions of masses, redshifts, and spins for all the sources are reported in Fig.~\ref{fig:universal_harmony_corner}. 

\begin{figure*}[t]
    \centering
    \includegraphics[width=.48\linewidth]{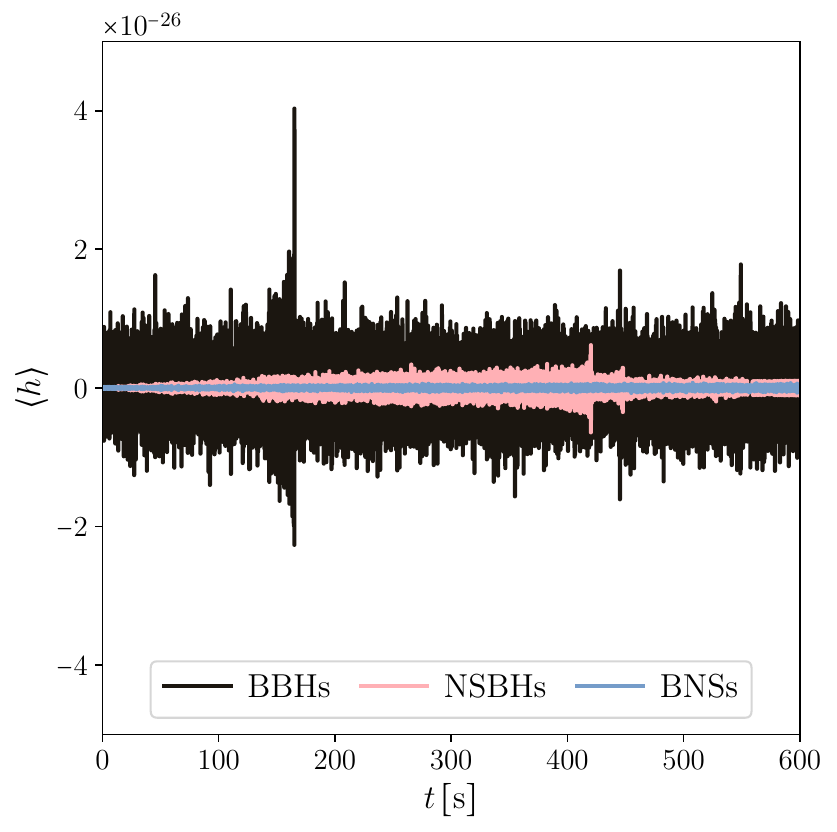} \hfill \includegraphics[width=.48\linewidth]{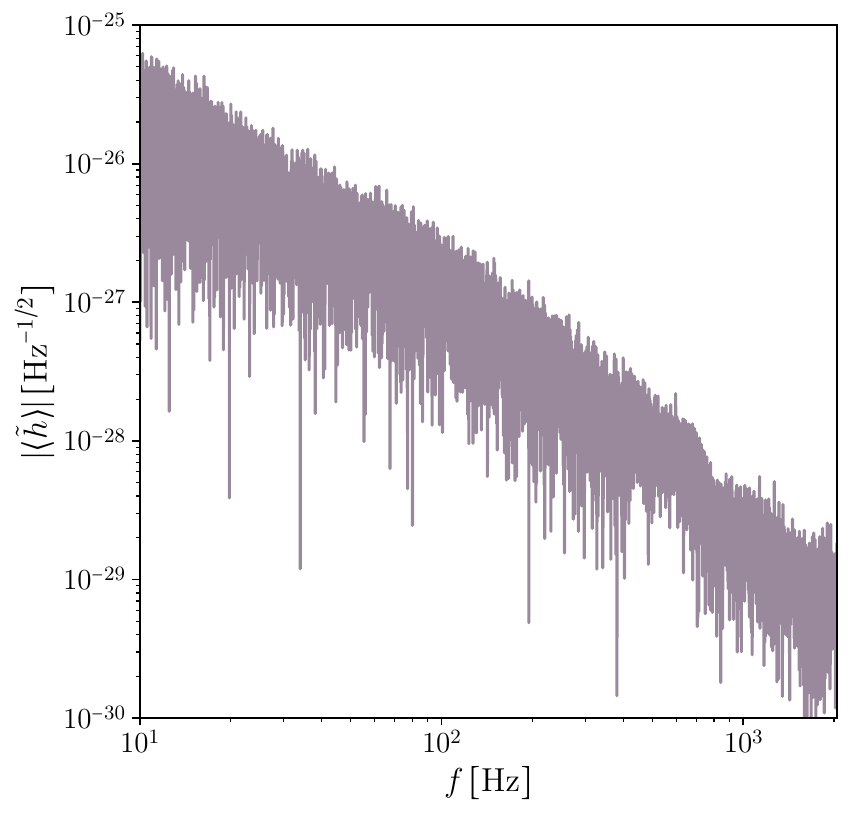}
    \caption{\emph{Left panel}: average 10\,min time-domain strain for BBH, BNS, and NSBH systems. \emph{Right panel}: total average frequency domain strain for the sum of the three considered populations of sources.}
    \label{fig:universal_harmony_strain}
\end{figure*}

For each source, I assign a time of merger from a uniform distribution between January 1\textsuperscript{st} 2026 and December 31\textsuperscript{st} 2026. I simulate the time-domain signal of each system in the synthetic catalogs using the \texttt{IMRPhenomT} waveform approximant~\cite{Estelles:2020osj} for BNSs and the \texttt{IMRPhenomTHM} model~\cite{Estelles:2020twz} for BBHs and NSBHs, which includes the contribution of sub-dominant harmonics. Each signal is computed from a minimum frequency of 10\,Hz, for listeners with six-million-dollar hearing, and sampled at 4096\,Hz. I then select chunks of 10\,min for the year-long datastream, and average over each of them. This is somewhat analogous to how recording engineers compress the dynamic range of music, though here I am compressing a year of the Universe's greatest hits into digestible segments. 

The resulting time-domain strain for the three kinds of CBCs considered and the total frequency-domain strain are reported in Fig.~\ref{fig:universal_harmony_strain}.

One can appreciate how BBH signals, despite not being the most numerous, dominate the strain, followed by NSBH systems and then BNSs. This is the GW equivalent of the loudest instruments carrying a musical piece: BBHs are clearly the Universe's drummers, pounding away with the most energy. BNSs, while more numerous, are the delicate string section---present and contributing to the overall texture, but easily overwhelmed.

One can also see that, due to the power-law nature of the frequency domain GW strain, $|\tilde{h}|\propto f^{-7/6}$, the dominant frequencies in the spectrum are the lowest. The Universe, it turns out, prefers bass to treble---a passion play that unfolds in the low-frequency regime. To compare with electromagnetic telescopes, about 20 years ago a group of astronomers from Johns Hopkins University has determined that the average color of the Universe is \colorbox{black}{\textcolor{cosmiclatte}{\texttt{\#FFF8E7}}}, also known as ``Cosmic Latte''~\cite{cosmiclatte_website}, which sits not far from the middle of the visible spectrum.

I convert the total time-domain strain into an audio file, scaling the amplitude to make it audible, while preserving the frequency content.

It is worth noting several caveats to my analysis. First, I have not included other potential GW sources such as continuous waves from rapidly rotating neutron stars, or burst signals from core collapse supernovae. These would add additional layers to the cosmic soundtrack, the continuous waves being a persistent hum, and the supernovae occasional screams. Second, my choice of averaging windows (10 minutes) is somewhat arbitrary, though I found it provides a reasonable balance between temporal resolution and signal accumulation. Shorter windows sound more variable; longer windows more monotonous. Finally, I acknowledge that this ``average'' sound necessarily destroys information about the temporal clustering and correlation of events, which could in principle add some texture to the cosmic audio. However, for the purposes of producing pleasant listening material, this sacrifice seems justified.

\section{Results: The Universal Harmony}

The resulting audio file, thick as a brick of compressed spacetime data, can be accessed \href{https://francescoiacovelli.github.io/universal_harmony/}{here}.\footnote{\url{https://francescoiacovelli.github.io/universal_harmony/}.} I encourage readers to listen with quality headphones to fully appreciate the nuances of the cosmic soundscape. The sound can be described as a low, rumbling whoosh, echoes of spacetime curvature, with occasional subtle frequency modulations---not unlike distant ocean waves in a stormy sea, or perhaps the ambient noise of a jet engine.

Some might describe it as soothing; others as slightly ominous; a few might think their stew (or polenta) is ready. I prefer to think of it as majestic. It is the sound of spacetime itself rippling, of the Universe's most extreme objects performing their final dance before merging forever.

\section{Conclusions}

I hope you enjoy this sound: it is a great substitute for white noise, allowing you to sleep cradled by the Universe. Unlike artificial white noise generators, this is the real deal---actual spacetime distortions, averaged over a year of cosmic violence, brought down to human-audible amplitudes through the magic of numerical scaling.

I have proposed here a complete pipeline for converting catalogs of synthetic GW events into audio files, making the Universal harmony audible. While this analysis was performed with a specific set of population models, the methodology is general and can be applied to any catalog of CBC events, or indeed to other GW sources, should they prove numerous enough to contribute meaningfully to the average.

Future work could explore several exciting directions. One could produce ``regional'' sounds by restricting to different hemispheres or volumes of the Universe, effectively creating a spatial map of cosmic acoustics---tales from topographic oceans of spacetime curvature. Time-dependent analyses could reveal whether the Universe's sound evolves over cosmological timescales. Different population models could be compared to assess how uncertainties in our current understanding of CBC demographics affect the Universal soundtrack. And of course, with the next generation of GW detectors, Einstein Telescope~\cite{Punturo:2010zz,ET:2025xjr} and Cosmic Explorer~\cite{Reitze:2019iox,Evans:2023euw}, we will produce a higher-fidelity recording, capturing fainter events and extending to higher and lower masses.

I also note that the results have potential applications beyond astrophysics. The sound could be used in meditation apps, as background ambiance in planetariums, or as a conversation starter at physics department parties. I am currently in discussions with several record labels about distribution rights.

In conclusion, I have answered the fundamental question: what does the Universe sound like? The answer: like a low rumble. Not as melodious as one might hope, but definitely better than bagpipes.

\section*{Data Availability}

The audio file is publicly available at \href{https://github.com/francescoiacovelli/universal_harmony/raw/main/output_audio_universal_harmony_all.wav}{github.com/francescoiacovelli/universal\_harmony/raw/ main/output\_audio\_universal\_harmony\_all.wav} and on arXiv, licensed under Creative Commons. I encourage remixes, though I note that the Universe retains all original performance rights. The synthetic catalogs used in this analysis are available in the respective references cited. No actual compact objects were harmed in the making of this analysis. A simplified version of the analysis scripts is available on \raisebox{-1pt}{\href{https://github.com/FrancescoIacovelli/universal_harmony}{GitHub \includegraphics[width=10pt]{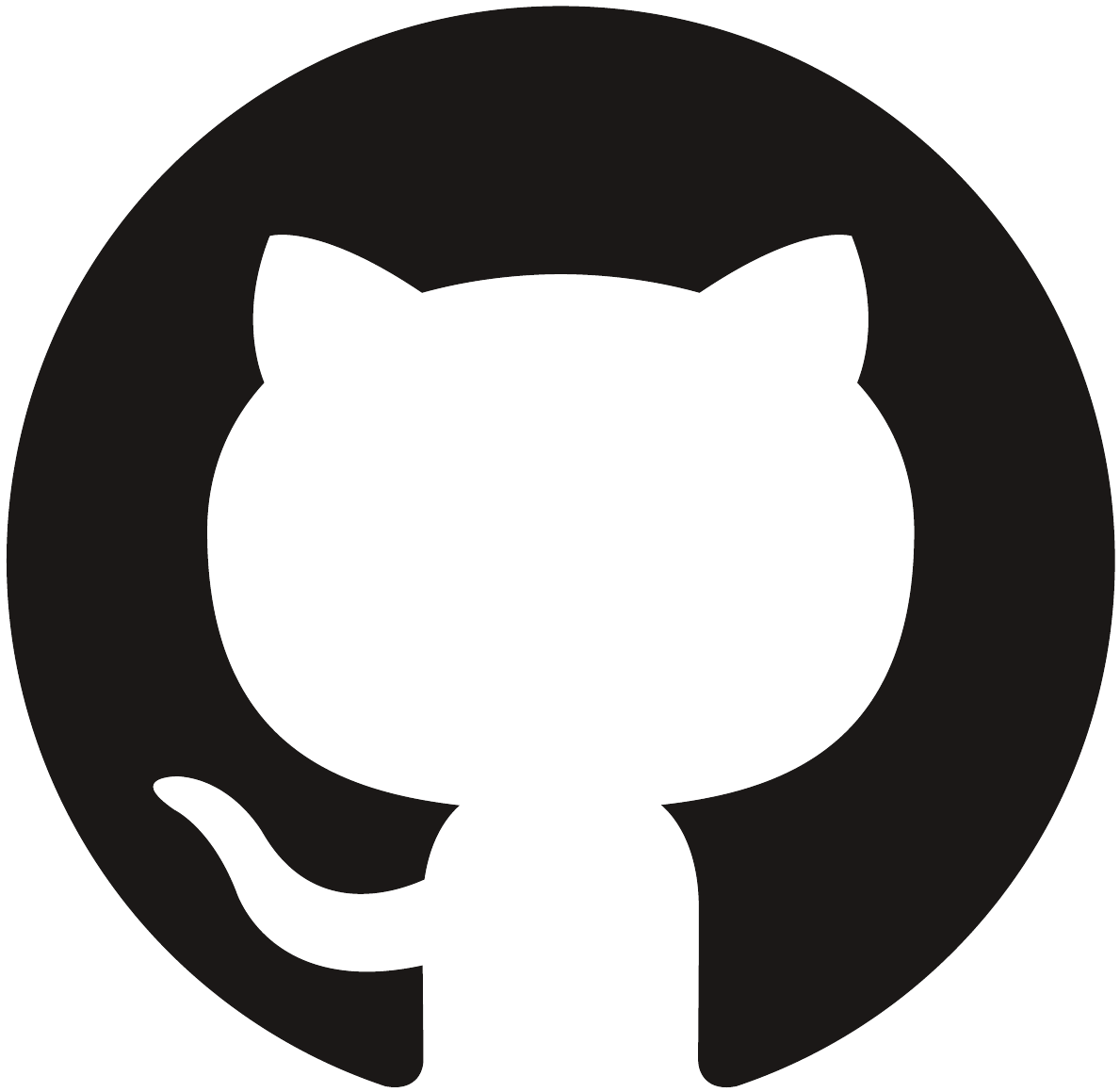}}}, allowing others to create their own Universal soundtracks, if desired.

\section*{Acknowledgments}

I am grateful to David Pereñiguez and Luca Reali for the initial spark of curiosity over coffee that led to this work, and for their subsequent decision to pursue other, presumably more serious, research directions. None of them could suggest how to make the Universe sound better. I thank Ulyana Dupletsa for listening to the audio file and confirming that yes, it is indeed there and sounds like a fireplace or polenta cooking, and for resisting the urge to question my life choices. Bogdan Ganchev deserves special recognition for actually listening to the full 10-minute audio file, an act of dedication that went above and beyond my expectations. 

\bibliography{universe_sound}

\end{document}